\documentstyle[preprint,tighten,aps,psfig]{revtex}

\begin{document}

\preprint{TTP97-05, hep-ph/9703291}
\draft

\date{March 1997}
\title{Two-loop QCD corrections to semileptonic $b$ decays at maximal
  recoil} 
\author{Andrzej Czarnecki and Kirill Melnikov}
\address{Institut f\"ur Theoretische Teilchenphysik, 
Universit\"at Karlsruhe,\\
D-76128 Karlsruhe, Germany}
\maketitle

\begin{abstract}
We present a complete ${\cal O}(\alpha_s^2)$  correction to the differential
width of the inclusive semileptonic decay $b\to cl\nu_l$ at the
kinematical point of vanishing invariant mass of the leptons, $q^2=0$.
Together with the recently computed  ${\cal O}(\alpha_s^2)$ correction 
at the upper boundary of the lepton invariant mass spectrum,
this new information permits an
estimate of the ${\cal O}(\alpha_s^2)$ effect in  the total inclusive
semileptonic decay width $b\to cl\nu_l$. We argue that the non-BLM part of
the ${\cal O}(\alpha_s^2)$ correction gives at most $1\%$ correction
to the inclusive semileptonic decay width $b\to cl\nu_l$. This significantly
improves the credibility of extracting $|V_{cb}|$ from the inclusive
semileptonic decays of the $b$-hadrons.
\end{abstract}

\vspace*{3mm}
Semileptonic decays of the $b$ quarks provide the best opportunity
to determine $|V_{cb}|$, a parameter of the Cabibbo-Kobayashi-Maskawa
(CKM) matrix and a fundamental input parameter of the standard model.
The current experimental limit \cite{RPP}
\begin{eqnarray}
|V_{cb}| = 0.036 \mbox{  to  } 0.046 \quad (90\% \mbox{ CL})
\end{eqnarray}
is based on measurements of the beauty hadron decays produced 
at the $\Upsilon(4S)$ resonance (by ARGUS
and by CLEO II) and in $Z$-boson decays  (by the four experiments at
LEP).  In the future large  samples of the $b$-hadrons collected at
$B$-factories (at SLAC and KEK) and at the hadron colliders will
increase the statistical accuracy to a few percent level.  To fully
exploit the anticipated experimental improvement, the theoretical
description of the $b$ decay must be known with comparable precision.

There are two methods of extracting the value of $|V_{cb}|$, based on
measurements of the exclusive decay $B\to {\bar D^\star} \bar l\nu_l$
and of the inclusive semileptonic decay width of $b$-hadrons $\Gamma
_{\rm sl}$.  These two methods rely on very different theoretical
considerations and experimental procedures and complement each other.
Their merits and theoretical uncertainties are summarized e.g.~in
Ref.~\cite{Shifman94,Neubert97,Bigi97}.  One of the major sources of the
theoretical error 
are the
perturbative QCD corrections at the two loop level.  For the exclusive
decays at the zero recoil point these corrections have recently been
calculated \cite{zerorecoil}.  This has significantly improved the
accuracy of the theoretical prediction for the exclusive method.

In the case of the inclusive semileptonic decay width of the
$b$-hadrons $\Gamma _{\rm sl}$, the only known effects beyond one loop
are those
associated with the running of the strong coupling
constant  \cite{Luke:1995,Ball:1995}.
They are obtained by computing massless quark effects
(fig.~1) 
and then replacing the number of light flavors $N_L$ by the combination
in which it enters the one-loop $\beta$-function $N_L - 33/2$.
These so-called
BLM corrections \cite{BLM} are expected to dominate the two-loop
result, however only a full calculation of the remaining diagrams will put
this statement on a firm foundation. 

Technically, the correction to the semileptonic decay width 
$\Gamma _{\rm sl}$ is obtained by fixing the invariant mass of the leptons
$q^2$ and computing the differential width  
${\rm d}\Gamma_{\rm sl} / {\rm d}q^2$ with desired
accuracy.  Integrating over $q^2$ within kinematical boundaries, 
one gets the inclusive
semileptonic decay width of $b \to c l\nu _l$:
\begin{eqnarray}
\Gamma _{\rm sl} = \int \limits _{0}^{(m_b-m_c)^2} {\rm d}q^2 ~
 \frac {{\rm d}\Gamma _{\rm sl}}{{\rm d}q^2}.
\label {width}
\end{eqnarray}

Going beyond the BLM approximation and computing complete
${\cal O}(\alpha_s^2)$ corrections
remains a daunting task at
present. In comparison with the zero recoil calculation 
the main difficulties are: an additional kinematical
variable describing the invariant mass of the leptons ($q^2$) and the
presence of the real radiation of one and two gluons.  

To circumvent these difficulties, we propose to estimate the
deviations from the BLM 
predictions by performing complete ${\cal
O}(\alpha_s^2)$ calcualtions for 
${\rm d}\Gamma_{\rm sl} / {\rm d}q^2$
at two boundaries of integration in 
eq.~(\ref {width}).

In fact, one of these calculations has already been done in 
ref. \cite {zerorecoil} where
${\cal O}(\alpha_s^2)$
corrections  to the transition $b \to c l \nu _l$ were
calculated at the zero recoil limit.
Since in this limit the radiation of real gluons is absent, the
results of \cite {zerorecoil} provide ${\cal O}(\alpha_s^2)$
correction to ${\rm d}\Gamma_{\rm sl} / {\rm d}q^2$ at 
$q^2_{\rm max} = (m_b-m_c)^2$.

The purpose of this paper is to present a calculation of the 
${\cal O}(\alpha_s^2)$ corrections at $q^2_{\rm min}=0$ which is the other
boundary for the invariant mass of the leptons. With both boundary
points known we can estimate the deviation of the ${\cal O}(\alpha_s^2)$
corrections to the total inclusive semileptonic decay width of the
$b$-quark $\Gamma _{\rm sl}$ from the BLM prediction.

Taking $q^2=0$ limit is important for the feasibility of this calculation. 
In this case the calculation 
of real radiation of one and two gluons is considerably simplified.

The reason why the real radiation at order ${\cal O}(\alpha^2_s)$ is
difficult to calculate is that the particle in the initial state (the
decaying $b$ quark) carries a color charge and therefore can radiate.
It is the presence of the
massive propagators of this particle which makes the integrations over
the phase space very tough.  For this reason even the QED corrections
to such well studied processes as the muon decay remain 
unknown at the two-loop level.  
The kinematical configuration in which $q^2=0$ and the quark
in the final state is massive is the first
case where the complete evaluation of the real radiation in the decay
of a fermion turns out possible.
Below we sketch the basic ideas of our approach; the technical details
will be presented elsewhere.

The idea which permitted us to calculate the contribution due to the
real radiation of one and two gluons is (qualitatively speaking) 
the expansion in
the velocity of the final quark. Indeed,
in the limit $m_c \to m_b$ the charm quark in the final state is a
slowly moving particle, with spatial components of its 
momentum of the order of 
$m_b-m_c$, much smaller than its mass.
The four momenta of gluons and of leptons (for $q^2 = 0$) 
are also of the order of $m_b - m_c$. It turns out that by a proper
choice of the phase space variables  one can systematically expand the  
amplitudes and the phase space in terms of $\delta \equiv (m_b-m_c)/m_b \ll 1$.

Some examples of the diagrams which contribute to the QCD corrections
to the semileptonic decay of the $b$ quark are shown in fig.~2. Not
shown are several other virtual corrections as well as diagrams 
obtained by permuting the gluon couplings
to the quark line or by crossing the external gluon lines.  In total
there are about 80 Feynman diagrams which have to be evaluated.

We do not include the diagrams with three $c$-quarks in the
final state in our analysis. Since 
$3m_c$ is only marginally smaller than $m_b$, the 
contribution of such diagrams is strongly suppressed.

We parametrize the expansion using the variable $\delta=1-m_c/m_b$.
In the first two nonvanishing orders ($\delta^3$ and $\delta^4$) only
virtual corrections contribute (e.g.~fig.~2a,b).  The following two
terms receive in addition contributions from diagrams with one loop
and one real gluon emission (like in fig.~2c,d), as well as from
diagrams with two gluons resulting from a decay of a virtual gluon
(fig.~2f).  Only in the order $\delta^7$ the contributions of a double
gluon emission from the quark line show up (fig.~2e). This hierarchy
can be traced back to the fact (evident in physical gauges) that
the interaction of the slowly moving quarks with real gluons 
is proportional to the three velocity of the former. 

In case of two-loop virtual corrections as well as in the emission of
two real gluons the expansion in $\delta$ means a
Taylor expansion in the small external momenta of
the leptons and gluons.  Such an expansion does not lead to any
spurious ultraviolet or infrared divergences.  The situation is
different in the case of the single gluon radiation in diagrams where
there is in addition one virtual loop (fig.~2c,d).  There a naive
Taylor expansion in the external gluon momentum leads to artificial
infrared divergences which correspond to the on-shell logarithmic
singularities of the one-loop diagrams. Therefore a more sophisticated
approach is needed and the recently developed method of ``eikonal
expansions'' \cite{Smirnov96,CzarSmir96} is used.

To present our result we write 
the differential semileptonic decay width of the decay $b\to c l \nu$
at $q^2=0$ as
\begin{eqnarray}
\Big [ \frac {{\rm d}\Gamma _{\rm sl}}{{\rm d} q^2} \Big ]_{q^2=0} = \Gamma_0
\left[
\Delta_{\rm Born}
+ {\alpha_s\over\pi} C_F \Delta_1
+ \left({\alpha_s\over\pi}\right)^2 C_F \Delta_2
\right]
\end{eqnarray}
where $\Gamma_0={G_F^2 m_b^3\over 96\pi^3}|V_{cb}|^2$ 
and $\Delta_{{\rm
    Born},1,2}$ describe the $m_c/m_b$ dependence in various orders in the
strong coupling constant.   

Both $\Delta_{{\rm Born}}=\left(1-m_c^2/m_b^2\right)^3$ 
and $\Delta _1$ are known in a closed analytical
form \cite{jk2,sa92b}.
$\Delta_2$ is the main result of the
present paper. For the purpose of  presentation we divide it up into
four contributions according to the color factors:
\begin{eqnarray}
\Delta_2 &=& \delta^3~~\left[ 
(C_F-C_A/2)\Delta_{F}
+C_A\Delta_{A}
+T_RN_L \Delta_{L}
+T_R \Delta_{H}
\right]
\end{eqnarray}
The last term, $\Delta_{H}$,
describes the contributions of the massive $b$ and $c$ quark loops.
Top quark contribution is suppressed by a factor $\sim m_b^2/m_t^2$
and has been neglected.

For the SU(3) group the color factors are $C_A=3$, $C_F=4/3$,
$T_R=1/2$.  $N_L=3$ is the number of the quark flavors whose masses
have been neglected ($u$, $d$, and $s$). 

We have computed the expansion coefficients of $\Delta_{F,A,L,H}$ up
to $\delta^8$, which for the physical value of the charm and bottom
masses gives an estimated 
accuracy of our numerical predictions better than 1\% (for 
$\delta=1-m_c/m_b \approx 0.7$).

In the present paper we list the analytical
results only up to $\delta^4$, while the numerical evaluation is done
using the expansions up to $\delta^8$.  Using the pole mass of the $b$
and $c$ quarks and expressing the one-loop corrections in terms of
$\alpha_{\overline{MS}}(m_b^2)$ we find
\begin{eqnarray}
\Delta_{A}  &=& 
       - {355 \over 36} + {2 \over 3} \pi^2
       + \delta   \left( {89 \over 8} - \pi^2 \right)
\nonumber\\ &&
       + \delta^2   \left(  - {2422517 \over 32400} + {1708 \over 45}
       \ln(2\delta) - {44 \over 9}  
         \ln^2(2\delta) + {8 \over 9} c_1 + {257 \over 90} \pi^2 \right)
\nonumber\\ &&
       + \delta^3   \left( {2956607 \over 64800} - {854 \over 45}
       \ln(2\delta) + {22 \over 9}  
         \ln^2(2\delta) - {4 \over 9} c_1 - {307 \over 180} \pi^2 \right)
\nonumber\\ &&
       + \delta^4   \left(  - {5789957 \over 1323000} + {4663 \over
         4725} \ln(2\delta) + {2 \over 5}  
         \ln^2(2\delta) + {4 \over 45} c_1 + {412 \over 1575} \pi^2 \right)
\nonumber\\
\Delta_{F}  &=&
       - {23 \over 6} + {8 \over 3} c_2 + {8 \over 3} \pi^2
       + \delta   \left( {23 \over 4} - 4 c_2 - 4 \pi^2 \right)
       + \delta^2   \left( {1697 \over 360} - {8 \over 3} \ln(2\delta)
       + {22 \over 5} c_2 + {359 \over 135} \pi^2 
          \right)
\nonumber\\ &&
       + \delta^3   \left(  - {3347 \over 720} + {4 \over 3}
       \ln(2\delta) - {23 \over 15} c_2 - {179 \over 270}  
         \pi^2 \right)
\nonumber\\ &&
       + \delta^4   \left( {4957991 \over 396900} - {1460 \over 189}
       \ln(2\delta) + {16 \over 9}  
         \ln^2(2\delta) + {2 \over 7} c_2 - {139 \over 600} \pi^2 \right)
\quad
\nonumber\\
\Delta_{L}  &=&       
        {14 \over 9}
       - \delta 
       + \delta^2   \left( {82217 \over 4050} - {544 \over 45}
       \ln(2\delta) + {16 \over 9} \ln^2(2\delta) 
          - {16 \over 27} \pi^2 \right)
\nonumber\\ &&
\quad
       + \delta^3   \left(  - {103667 \over 8100} + {272 \over 45}
       \ln(2\delta) - {8 \over 9}  
         \ln^2(2\delta) + {8 \over 27} \pi^2 \right)
\nonumber\\ &&
       + \delta^4   \left( {1322183 \over 496125} - {2404 \over 1575}
       \ln(2\delta) + {8 \over 45}  
         \ln^2(2\delta) - {8 \over 135} \pi^2 \right)
\nonumber\\
\Delta_{H}  &=&
        {460 \over 9} - {16 \over 3} \pi^2
       + \delta   \left(  - 74 + 8 \pi^2 \right)
       + \delta^2   \left( {9821 \over 81} - {344 \over 27} \pi^2 \right)
\nonumber\\ &&
       + \delta^3   \left(  - {33883 \over 810} - {32 \over 9}
       \ln(2\delta) + {136 \over 27} \pi^2 \right) 
       + \delta^4   \left( {3754 \over 405} - {154 \over 135} \pi^2 \right)
\nonumber \\
\end{eqnarray}
with $c_1={21\over 2}\zeta_3-\pi^2\ln(2\delta)$
and $c_2={3\over 2}\zeta_3-\pi^2\ln 2$.

We now turn to the numerical analysis of our result. 
Here the issue of numerical values for the quark masses becomes important.
It is  safe to assume that the pole mass of the $b$--quark lies 
between $4.6$ and $5.1$ GeV. 
The mass of the $c$ quark is determined by $m_b-m_c$, obtained from
the HQET calculations \cite {Shifman94,Neubert97,Ball:1995,Bigi97}.
We use $m_b-m_c \approx 3.45 \pm 0.10$ GeV where the 
error bar is rather conservative.

Accordingly, the numerical value of $\delta$ changes within the range 
of $0.65-0.77$. The numerical values for the function  $\Delta_2$ 
become:
\begin{eqnarray}
\Delta _2 = -6.03,~~-7.45(4),~~-8.96, 
\label {number}
\end{eqnarray}
for $\delta = 0.65,~~0.7,~~0.75$ respectively.

The error estimate, shown for the central value of $\delta = 0.7$, 
is obtained by multiplying the last computed term
by 3, which corresponds roughly to summing up the remainder of the
series in $\delta$ assuming constant coefficients.  This procedure
overestimates the error because the coefficients in fact decrease
(there is at most a logarithmic divergence at $\delta=1$ caused by
neglected diagrams with three real $c$ quarks in the final state). 

Taken literally, the ${\cal O}(\alpha_s^2)$ corrections are quite large.
However, as we will show below, the bulk of them  is
due to the BLM corrections.

The BLM prediction with 4 light
flavors of quarks gives the following results:
\begin{eqnarray}
\Delta_2^{\rm BLM} = -\delta^3 \Delta_L T_R\left(
{33\over 2}-4
\right)
=-6.54,~~~-8.15(6),~~~-9.87,
\label {numberBLM}
\end{eqnarray}
for $\delta = 0.65,~~0.7,~~0.75$ respectively.

By comparing the numbers in eq.~(\ref {numberBLM}) with those in
eq.~(\ref {number}) 
we conclude that the BLM correction accounts for most of the effect.
We estimate the residual correction by subtracting the BLM
piece from the exact correction. We get
a residual correction $(0.51,~0.7,~0.91)C_F(\alpha_s/\pi)^2$, which,
using  $\alpha _s(m_b) = 0.23$, gives 
numerically $0.5,~~0.7,~~0.8 \%$  correction relative to the Born rate for 
$\delta = 0.65,~~0.7,~~0.75$.

Therefore, we arrive at the conclusion that at the lower boundary
of the invariant masses of leptons  $q^2 _{\rm min} = 0$, 
the BLM piece of the ${\cal O} (\alpha _s ^2)$
correction represents the complete result 
with an excellent accuracy.
The remaining correction does not exceed the value of $1 \%$ even
accounting for an uncertainty in input parameters.

Finally, we would like to estimate the ${\cal O} (\alpha _s ^2)$ 
radiative corrections to the total semileptonic decay width of the $b$ quark. 
In the BLM approximation such corrections have been calculated in
ref.~\cite{Luke:1995,Ball:1995}. Therefore, we are only interested in
the deviations from  the BLM approximation.  

Our estimate of the non-BLM corrections to the inclusive
width  is based on the expectation
that the largest deviation from  BLM should 
occur at the maximal recoil limit, i.e. at $q^2=0$. 
To clarify this point, we note that the results of ref.~\cite
{zerorecoil} imply that at zero recoil limit ($q^2_{\rm max} = (m_b-m_c)^2$)
the deviation of the exact result from the BLM approximation are
very small. On the other hand, the results of this paper show that
at $q^2=0$ the non-BLM part of the correction grows with the  decrease of the
$c$-quark mass, i.e. with the increase in the phase space available 
for real gluon radiation. If one fixes the value of the $c$-quark
mass, but varies instead the invariant mass of leptons $q^2$, the
strongest emission of real gluons will occur at the maximal recoil
point of the spectrum, at $q^2=0$. It is for this reason that 
we expect the largest discrepancy between the BLM prediction and the
full correction at the lower end of the $q^2$ distribution, for
$q^2=0$. 

Turning to the estimate itself, from ref.~\cite{zerorecoil} we know
that at $q^2_{\rm max} = (m_b-m_c)^2$ (zero recoil limit) the non-BLM
correction to the differential width relative to the Born value is of
the order of $-0.1\%$. On the other end of the lepton invariant mass
distribution the result of this paper implies a slightly larger, but
also tiny deviation below $1\%$. We note that the change of sign of
the non-BLM corrections cancels part of their impact on the total
width.  
Taking the absolute value of the
larger of the corrections at the boundaries
as an upper bound 
we conclude
{\it that the non-{\rm BLM} piece of the ${\cal O} (\alpha _s ^2)$
corrections to the total semileptonic decay width $b \to c l \nu _l$
should not exceed the value of $1\%$}.

The value of second order correction to the inclusive width depends on
the adopted definition of the quark masses. 
Our result is presented in terms
of the pole masses, which is a convenient choice for the corrections
not associated with the running of the coupling constant.  It was
argued in \cite {Vai97} that such parametrization leads to small
higher--order non--BLM corrections.  Our result confirms this
expectation.

It is fair to say at this point that our estimate of the non--BLM
piece of the corrections to the total inclusive semileptonic decay
width based on the two boundary values can not be considered as a rigorous
proof. Keeping in mind that the complete calculation of the two loop
QCD corrections to the total decay width remains a very
difficult task, a calculation of these corrections at some
intermediate point $q^2_{\rm int}$ for the differential 
inclusive semileptonic decay width of the $b$--quark is highly desirable.
If such a calculation confirms that the non BLM
piece of the correction remains within the range set by its value on
two boundaries, our estimate for the correction to the total
semileptonic decay width of the $b$ quark will be on a very safe ground. 

We are grateful to K.~G.~Chetyrkin, M.~Neubert and N.~G.~Uraltsev 
for discussions and advice in the course of this calculation.
We would like to thank Prof. J.~H.~K\"uhn for his interest in this
work and support. 
This research has been supported by BMBF 057KA92P  and by
Graduiertenkolleg ``Teilchenphysik'' at the University of Karlsruhe.

\begin{figure} 
\hspace*{-36mm}
\begin{minipage}{16.cm}
\vspace*{5mm}
\[
\mbox{
\begin{tabular}{cc}
\psfig{figure=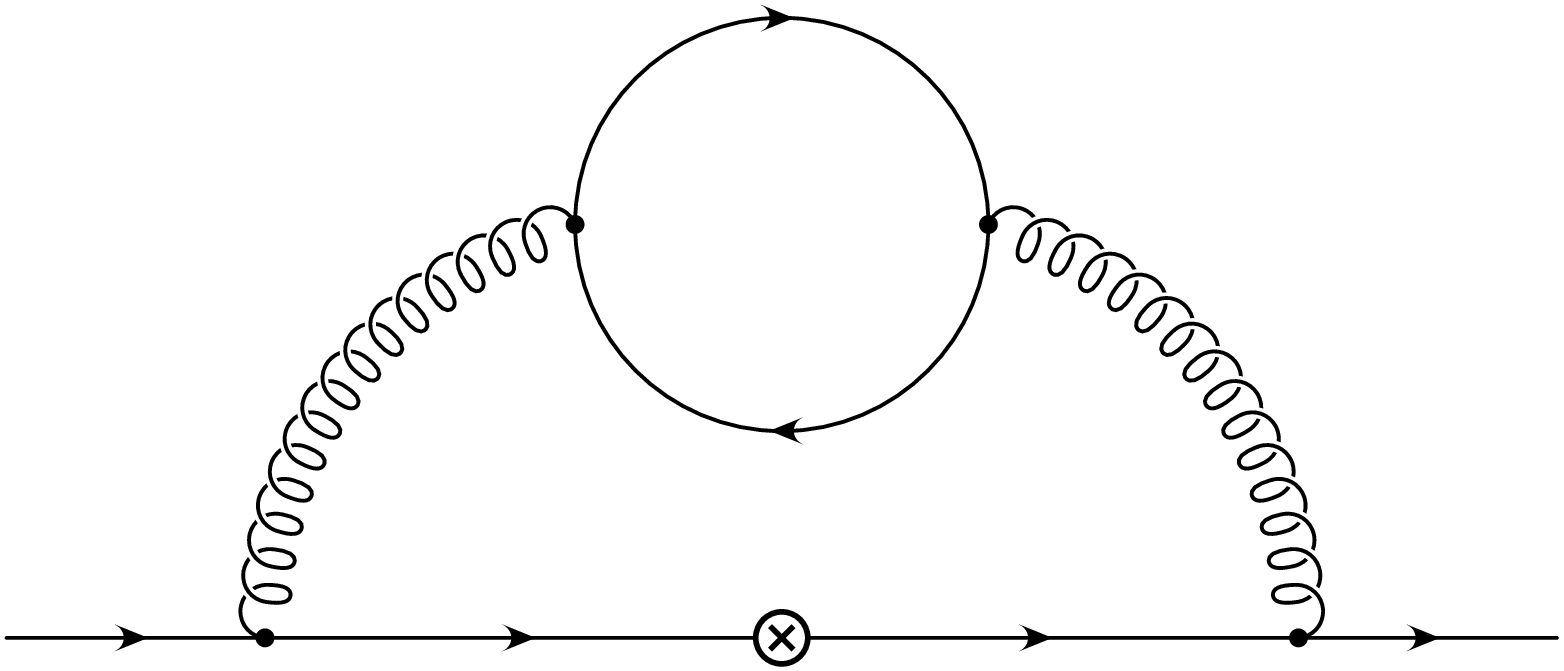,width=23mm,bbllx=210pt,bblly=410pt,%
bburx=630pt,bbury=550pt} 
&\hspace*{19mm}
\psfig{figure=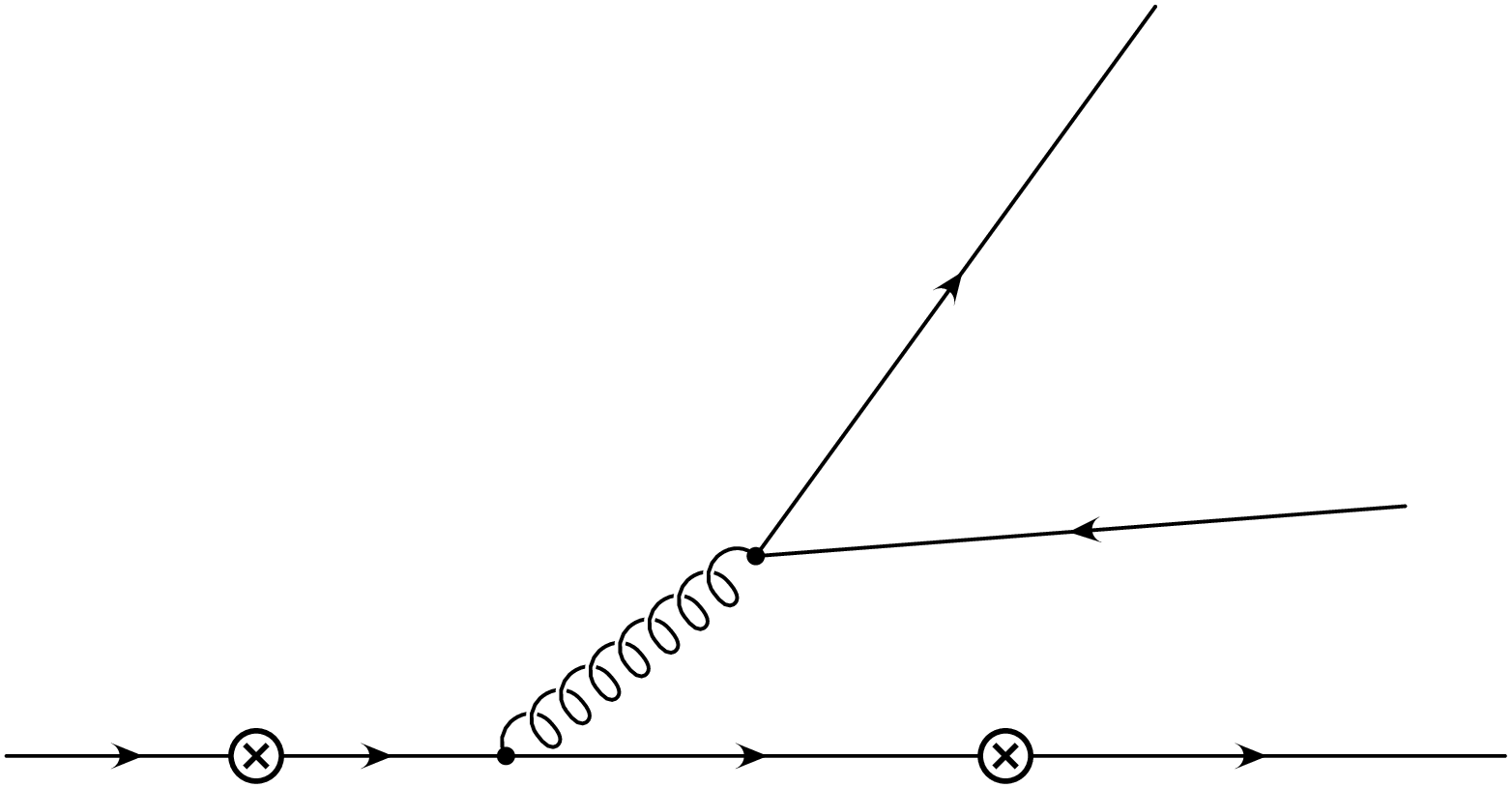,width=23mm,bbllx=210pt,bblly=410pt,%
bburx=630pt,bbury=550pt}
\\[5mm]
\hspace*{-13mm}(a) & \hspace*{7mm}(b) 
\end{tabular}}
\]
\end{minipage}
\caption{
Diagrams involving a light quark loop (a) or real pair emission (b).
Symbols $\otimes$ mark places where the virtual $W$ boson can
possibly couple to the quark line.}
\label{fig:BLM}
\end{figure}

\begin{figure} 
\hspace*{-36mm}
\begin{minipage}{16.cm}
\vspace*{5mm}
\[
\mbox{
\begin{tabular}{cc}
\psfig{figure=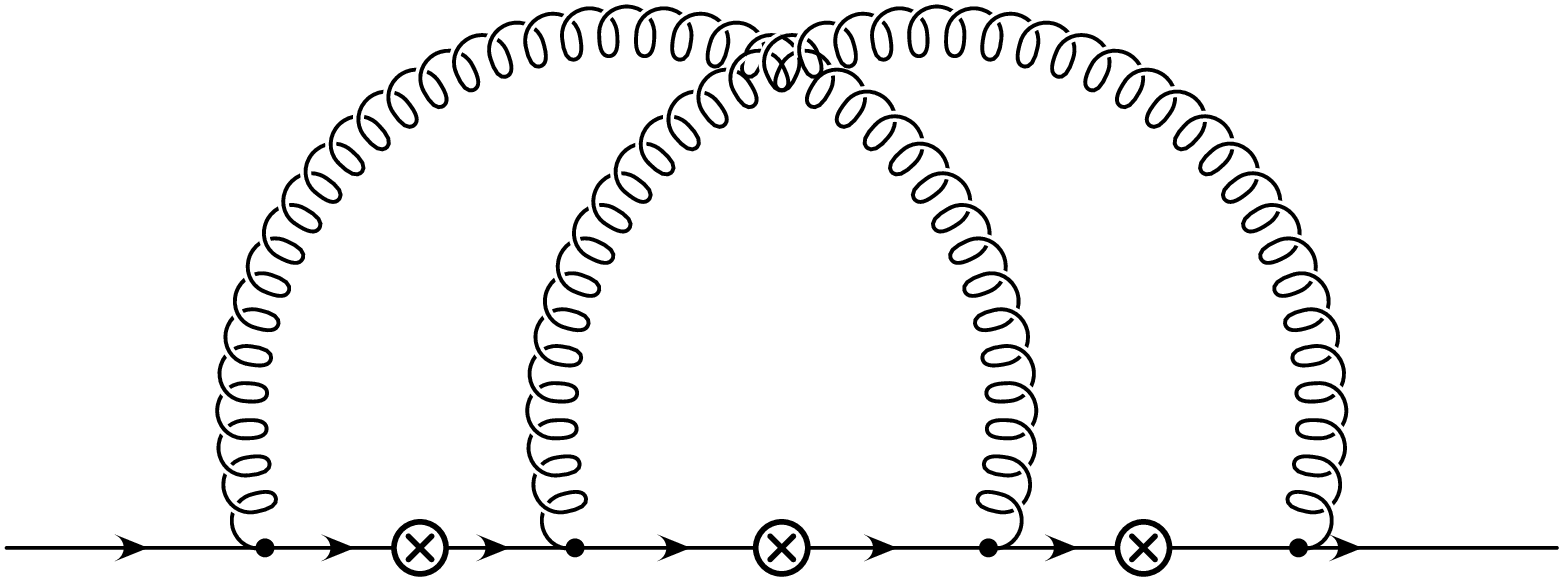,width=23mm,bbllx=210pt,bblly=410pt,%
bburx=630pt,bbury=550pt} 
&\hspace*{19mm}
\psfig{figure=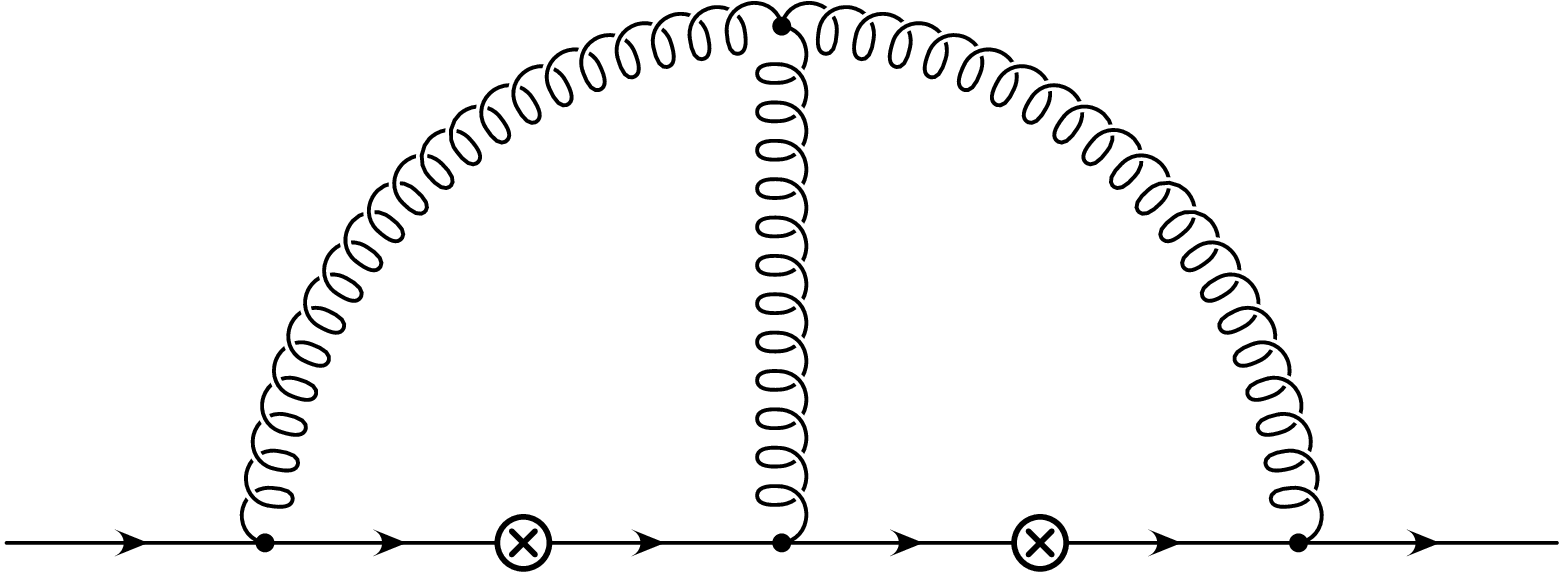,width=23mm,bbllx=210pt,bblly=410pt,%
bburx=630pt,bbury=550pt}
\\[5mm]
\hspace*{-13mm}(a) & \hspace*{7mm}(b) 
\\[12mm]
\psfig{figure=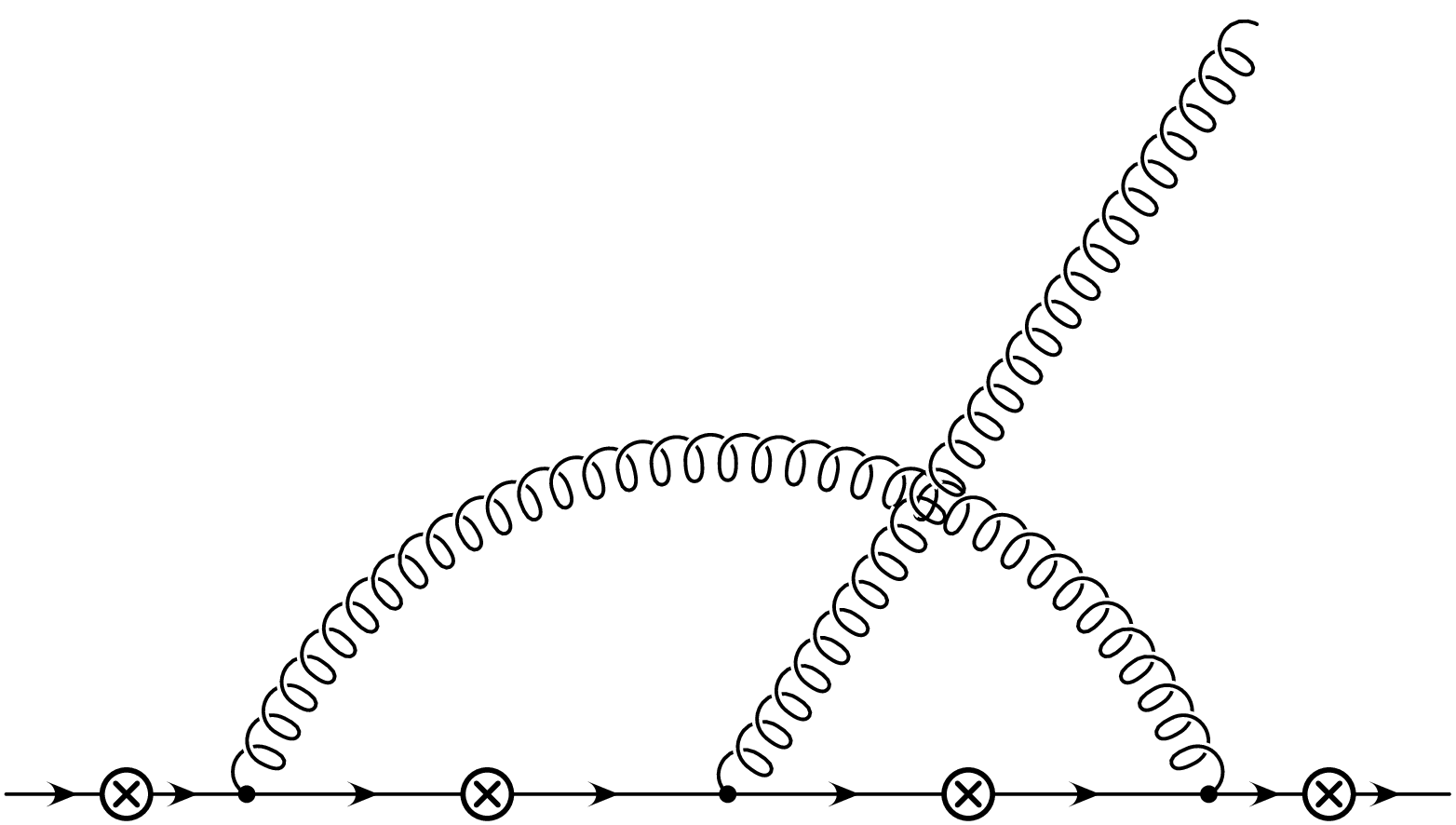,width=23mm,bbllx=210pt,bblly=410pt,%
bburx=630pt,bbury=550pt}
&\hspace*{19mm}
\psfig{figure=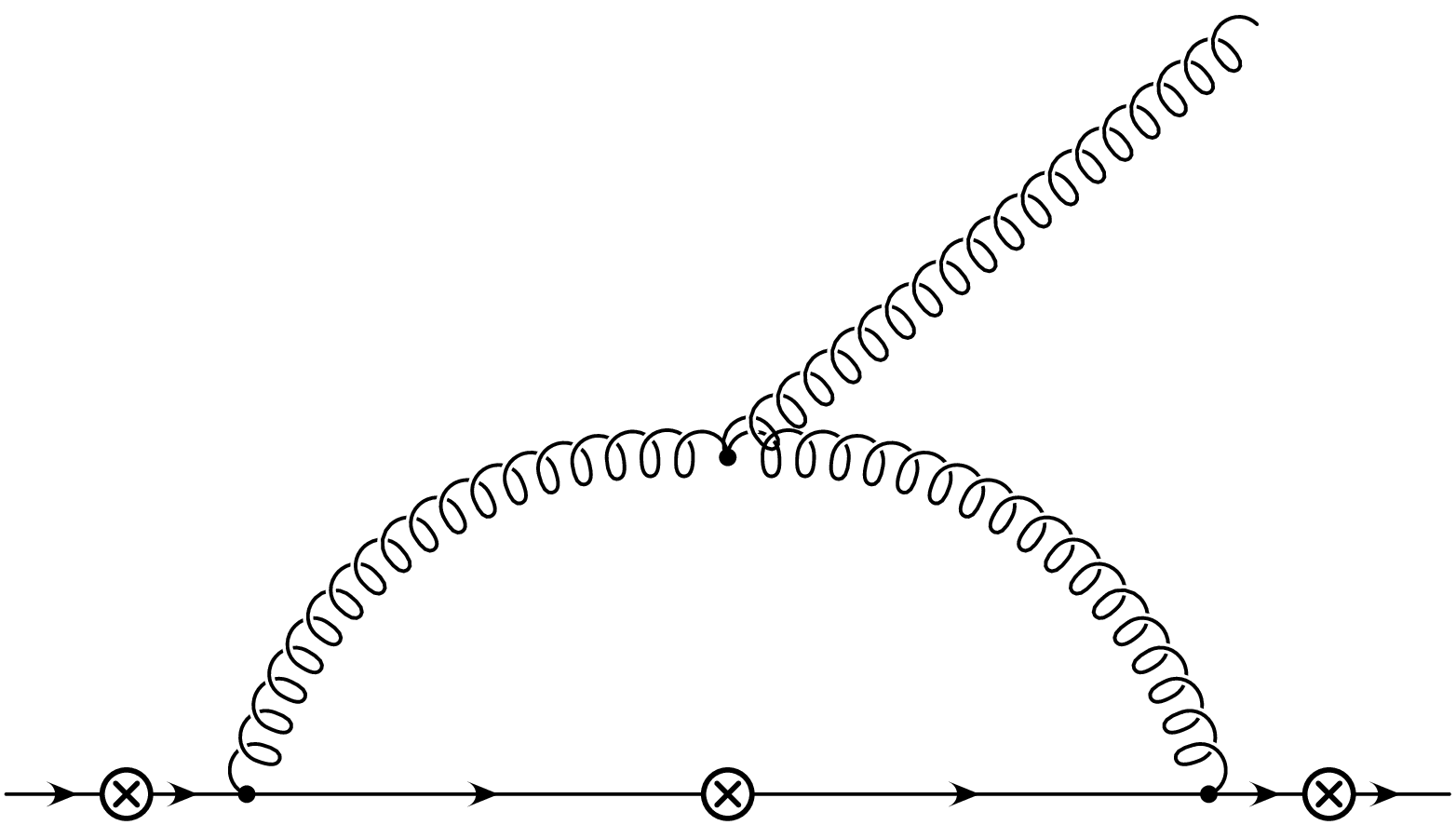,width=23mm,bbllx=210pt,bblly=410pt,%
bburx=630pt,bbury=550pt} 
\\[5mm]
\hspace*{-13mm}(c) & \hspace*{7mm}(d) 
\\[10mm]
\psfig{figure=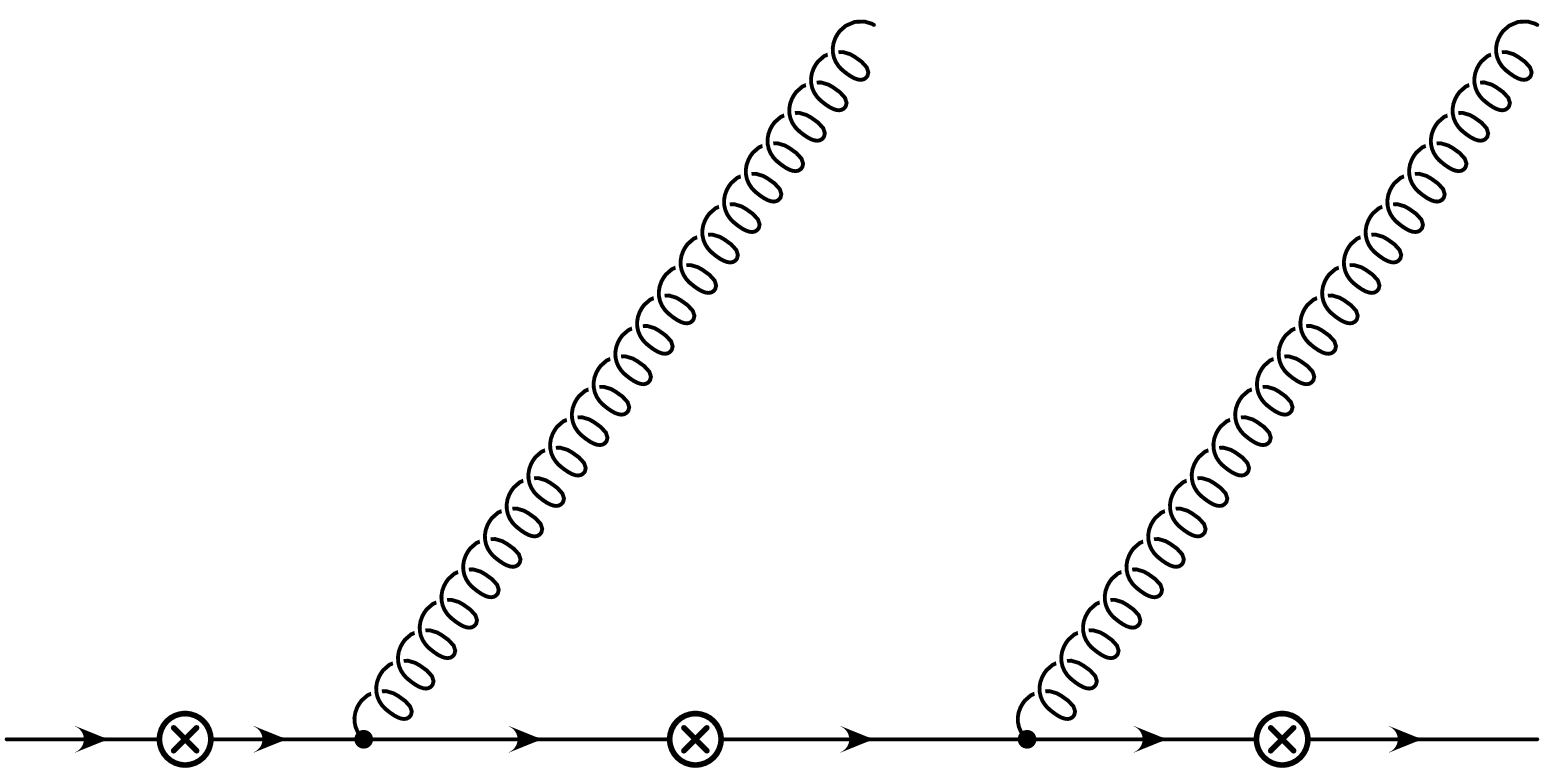,width=23mm,bbllx=210pt,bblly=410pt,%
bburx=630pt,bbury=550pt}
&\hspace*{19mm}
\psfig{figure=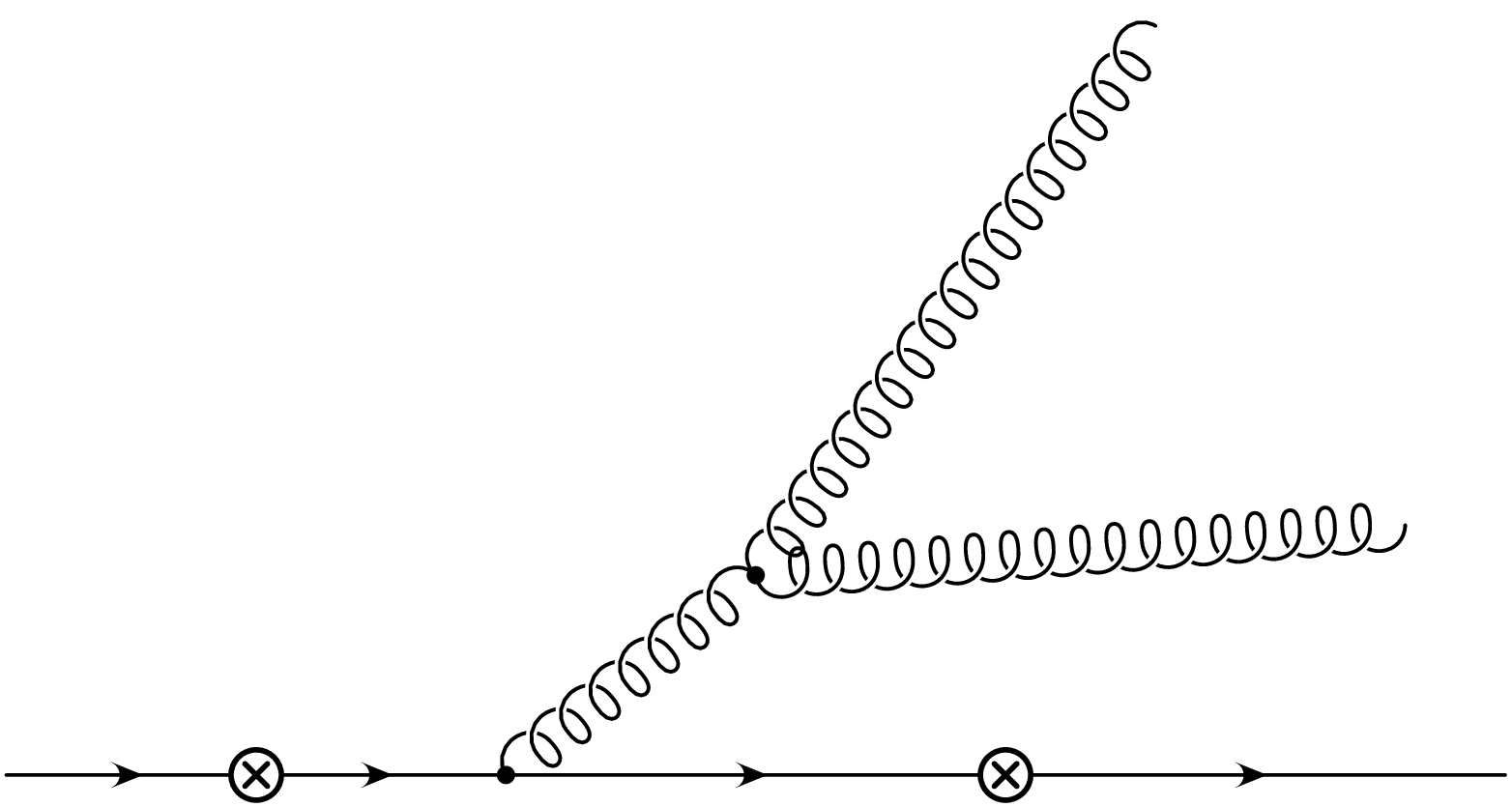,width=23mm,bbllx=210pt,bblly=410pt,%
bburx=630pt,bbury=550pt}
\\[4mm]
\hspace*{-13mm}(e) & \hspace*{7mm}(f) 
\end{tabular}}
\]
\end{minipage}
\caption{Examples of the 
two-loop gluonic QCD corrections to the decay $b\to c l\nu_l$  at zero
recoil; a,b: virtual corrections; c,d: single gluon emission; e,f:
emission of two
gluons.   Symbols $\otimes$ mark places where the virtual $W$ boson can
possibly couple to the quark line. The left hand side diagrams are
QED-like, while the right hand side ones are purely nonabelian. }
\label{fig:twoloop}
\end{figure}

\end{document}